# The James Webb Space Telescope's plan for operations and instrument capabilities for observations in the Solar System

Stefanie N. Milam[1], John A. Stansberry[2], George Sonneborn[1], Cristina Thomas[1,3,4]

[1]NASA Goddard Space Flight Center, 8800 Greenbelt Rd, Greenbelt, MD 20771
(stefanie.n.milam@nasa.gov; george.sonneborn-1@nasa.gov; cristina.a.thomas@nasa.gov)
[2]Space Telescope Science Institute, 3700 San Martin Drive, Baltimore, MD 21218 (jstans@stsci.edu)
[3]NASA Postdoctoral Program Fellow, NASA Goddard Space Flight Center, 8800 Greenbelt Rd, Greenbelt, MD 20771
[4]Planetary Science Institute, 1700 East Fort Lowell, Suite 106, Tucson, AZ 85719

**Abstract:**

The James Webb Space Telescope (JWST) is optimized for observations in the near and mid infrared and will provide essential observations for targets that cannot be conducted from the ground or other missions during its lifetime. The state-of-the-art science instruments, along with the telescope's moving target tracking, will enable the infrared study, with unprecedented detail, for nearly every object (Mars and beyond) in the solar system. The goals of this special issue are to stimulate discussion and encourage participation in JWST planning among members of the planetary science community. Key science goals for various targets, observing capabilities for JWST, and highlights for the complementary nature with other missions/observatories are described in this paper.

1. Introduction

The James Webb Space Telescope (JWST) is an infrared-optimized observatory with a 6.5m-diameter segmented primary mirror and instrumentation that provides wavelength coverage of 0.6 to 28.5 microns, sensitivity 10X to 100X greater than previous or current facilities, and high angular resolution (0.07 arcsec at 2 microns). Details on the science goals, mission implementation, and project overview are summarized in Gardner et al. (2006). The science instruments have imaging, coronagraphic, and spectroscopic modes that provide spectral resolving power of ~100 < R < ~3000, integral field units, and near-infrared multi-object spectroscopy. The capabilities of JWST will enable important studies throughout the Solar System beyond Earth's orbit, many of which are discussed in this special issue. JWST will provide access to the important 3 micron region free of the strong atmospheric absorptions that restrict observations from Earth. The Target of Opportunity response time can be as short as 48 hours, enabling timely observations of important events.

The JWST mission is led by NASA's Goddard Space Flight Center, with major mission participation, including science instrumentation, by the European Space Agency (ESA) and Canadian Space Agency (CSA). The JWST observatory will be launched by an Ariane 5 rocket provided by ESA. The observatory is designed for a 5-year prime science mission, with consumables for 10 years of science operations. JWST will be operated for NASA by the Space Telescope Science Institute in Baltimore, Maryland. The first call for proposals for JWST observations will be released in late 2017.

After launch in October 2018, JWST will enter a Lissajous orbit, free of Earth and Lunar eclipses, around the Sun-Earth L2 point. This orbit simplifies planning and scheduling, and minimizes thermal and scattered light influences from the Earth and Moon. The telescope and science instruments are passively cooled to 40K by remaining in the shadow of a 160 m$^2$ five-layer sunshield. The mid-IR detector is further cooled to 6.5K by a cryocooler.

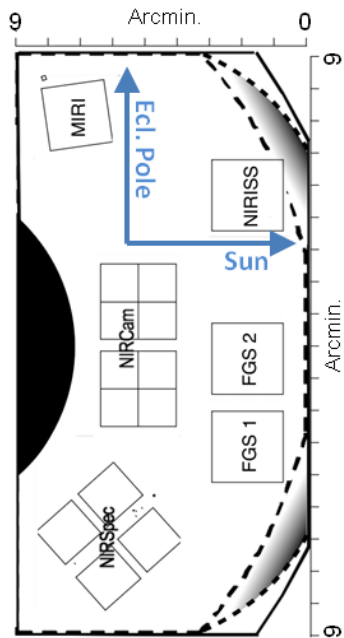

**Figure 1**. JWST science Instrument and guider fields of view as they project onto the sky. The Sun and ecliptic north directions are shown for a line-of-sight in the ecliptic plane, at a solar elongation of 90º, and in the direction of observatory orbital motion. **The line of sight is restricted to elongations 85º – 135º**.

The optics provide diffraction-limited performance at wavelengths ≥ 2µm (where the PSF FWHM is 64 mas). Scattered light is a concern for observing near bright targets such as the giant planets and brighter asteroids. Details of mirror-segment edges are being incorporated into JWST PSF models, and will improve their fidelity. Unfortunately accurate scattered light performance will not be known until on-orbit testing is completed. The commissioning plan for the observatory will include such testing, and the scope of the tests is under discussion as of this writing.

The four science instruments on JWST (NIRCam, NIRISS, NIPSpec and MIRI) cover the wavelength range from 0.6-28.5 microns, and offer superb imaging and spectroscopic sensitivity (see some additional detail in Tables 1 and 2 below). Full details on the MIRI instrument can be found in the special issue of PASP recently published (e.g. Rieke et al. 2015). Subarray readouts will enable observations of the giant planets and many bright primitive bodies in a variety of instrument modes. Saturation limits and special modes are presented in Norwood et al. (2015). The science instruments share the telescope focal plane with a Fine Guidance Sensor (FGS - a part of the CSA contribution to the mission) as illustrated in Figure 1.

Moving target observations are executed by controlling the position of a guide star in the FGS such that the science target remains stationary in the specified science instrument. The JWST attitude control system will track objects moving at rates of at least 30 mas/sec. The target ephemeris is represented in the observatory attitude control system as a 5th order polynomial, enabling tracking of objects (such as Io) that have large apparent accelerations. Pointing stability (and therefore image quality) for moving targets is predicted to be better than 10 mas over 1000 seconds, comparable to that for fixed targets.

Figure 2 illustrates how the JWST "speed limit" of 30 mas/sec affects the kinds of targets that can be observed, and at what heliocentric distances. Ephemerides for 11,000 near-Earth objects (NEOs), 170 known comets, 300 main-belt asteroids (MBAs), and 130 Centaurs and trans-Neptunian objects (TNOs) were retrieved from the Jet Propulsion Laboratory's Horizons system for the years 2019-2020 at one-day spacing. JWST was chosen as the observer location (specified by entering "@JWST" for the observatory), and the analysis was restricted to times when the target fell within the JWST field of regard (see below).

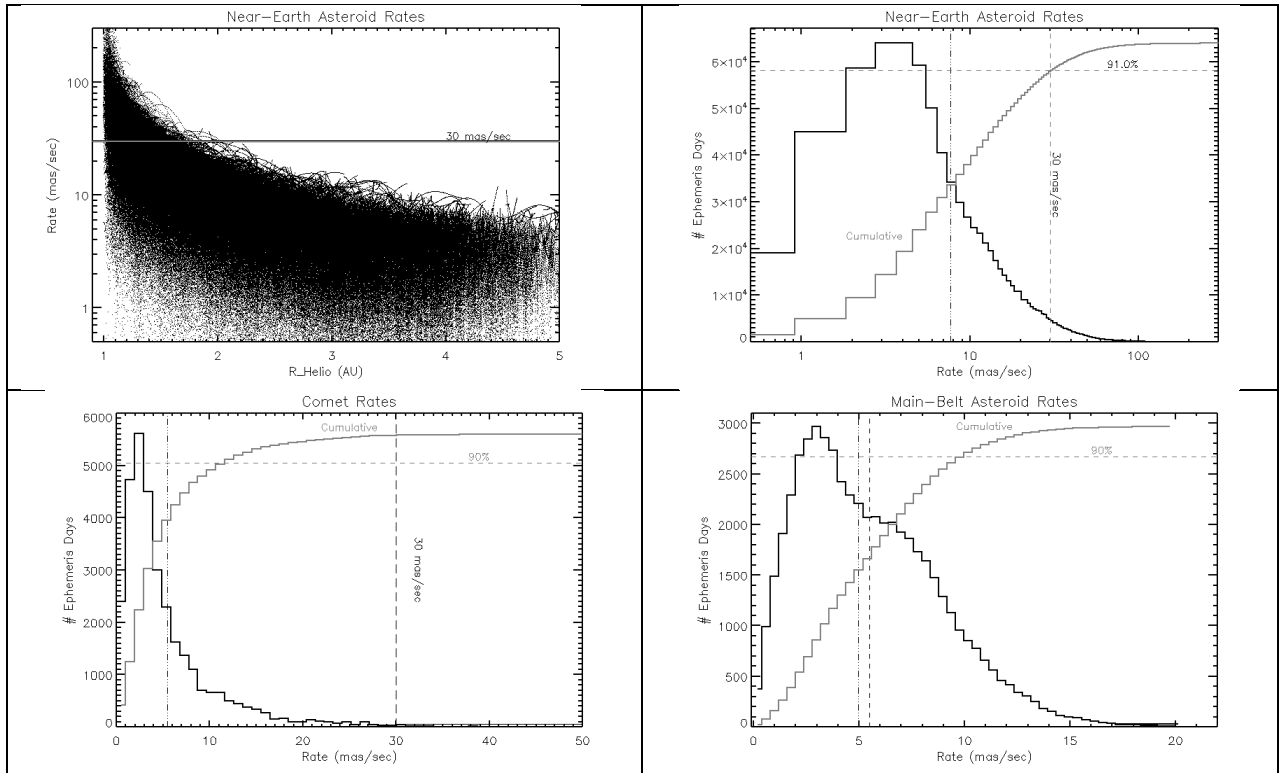

**Figure 2**. Apparent rates of motion of 11,467 Near Earth Objects (observable in 2019), 170 Comets, and 305 Main Belt Asteroids as seen from JWST during 2019 – 2020. Only the rates while objects are in the 85 – 135 deg elongation range are included. Top Left: Points each give the apparent rate for one object on one date, plotted vs. the heliocentric distance of the object on that date. Top Right: Differential (black line) and cumulative (gray line) histograms of the data shown in the top left panel. Vertical lines give the mean apparent rate for the sample (dash triple-dot) and the 30 mas/sec JWST speed limit (dashed). The cumulative curve is re-normalized to be on the same scale as the differential curve. On any given date when an NEO is within the JWST elongation limits, there is a 91% probability that it can be tracked at or below the 30 mas/sec limit. Bottom Left: Similar histograms for known comets. Only one comet can't be observed on any date in the two year period, while 6 can only be observed for a limited time. Bottom Right: Similar histograms, but for the MBA sample. While these results are generally encouraging, it is important to note that signal-to-noise in a given exposure time and spatial resolution on a target both peak at epochs when the target is closest to the observatory, and that the apparent rates of motion are typically highest at that time. If JWST can be made to track targets at rates higher than 30 mas/sec, there will be a significant benefit in reducing necessary integration times and enhanced resolution.

The biggest impact of the 30 mas/sec tracking-rate limit is for NEOs, where for any randomly chosen day there is a 9% chance that the target would be moving too fast to be tracked (see also Thomas et al., this issue). Panel (a) of Figure 2 shows that NEOs are preferentially moving too fast for JWST when they are nearest the Sun (this is primarily because that is when they are closest to JWST as well). For the known comets there is about a 1% probability that a target can't be tracked on any given day. For MBAs, Centaurs and TNOs, all targets can be tracked on any day. The JWST ephemeris currently available in Horizons is nominal, but representative of the class of orbits the observatory may ultimately follow. Plans are in place to update the observatory ephemeris in Horizons about every two weeks after launch, so uncertainties in the JWST position will be a small component in the uncertainty on moving target location in the science data. Absolute pointing accuracy is limited by the accuracy of the guide-star catalog and on the moving-target ephemeris itself. Accuracy of the guide-star catalog is expected to improve dramatically as a result of the GAIA mission, so JWST pointing accuracy relative to guide-stars should be well under 1". Ephemeris uncertainty depends on the particular science target, and is not particular to whether a target is observed from JWST or elsewhere.

The JWST sunshield design requires the orientation of the telescope line of sight to always be in the solar elongation range of 85° to 135° (see Figure 1). The observatory boresight can be pointed to any position on the celestial sphere within this elongation range. This fundamental pointing limitation for JWST is similar to that for other passively-cooled infrared missions such as *Spitzer* and *Herschel* (Wilson & Scott 2006; Pilbratt et al. 2010). Objects close to the ecliptic are typically observable during two ~50 day intervals per year. However, the motion of some Solar System objects may further limit this observability. For example, Mars is only observable in 2018, 2020, and 2022 (see Villanueva et al. this edition). When the observatory pointing lies near the ecliptic plane the instrument fields of view are highly restricted in their position angle regardless of date or solar elongation, as shown in Figure 1. The roll angle about the boresight has a +/- 5° range.

2. **Solar System Science**

"Planetary systems and the origin of life" is one of the core science themes for JWST. For example within the Solar System, molecular inventories and surface composition of Kuiper Belt Objects and comets will provide key insight into the dynamic history of the Solar System and help constrain current theories. Global scale imaging and spectroscopy of planetary atmospheres will be used to routinely measure temporal variations as well as decipher dynamics and chemistry. Norwood *et al.* (2015) updates Lunine et al. (2010) and provides an overview and some preliminary Solar System science case studies and we present top level highlights here. In addition, 10 community-based focus groups have thoroughly examined JWST's capabilities in the areas of: Near Earth Objects (Thomas et al.), Asteroids (Rivkin et al.), Comets (Kelley et al.), Giant Planets (Norwood et al.), Mars (Villanueva et al.), Occultations (Santos-Sanz et al.), Rings (Tiscareno et al.), Satellites (Keszthelyi et al.), Titan (Nixon et al.), and Trans-Neptunian Objects (Parker et al.). The JWST project has provided technical information and guidance to these groups to aid in their work. Detailed results from these studies are published in this special edition in the following articles:

- The James Webb Space Telescope's plan for operations and instrument capabilities for observations in the Solar System – S.N. Milam et al. (this article)
- Observing Near-Earth Objects with the James Webb Space Telescope – C.A. Thomas et al.
- Asteroids and JWST – A. Rivkin et al.
- Unique Spectroscopy and Imaging of Mars with JWST – G. Villanueva et al.
- Giant Planet Observations with the James Webb Space Telescope – J. Norwood et al.
- Observing Outer Planet Satellites (except Titan) with JWST: Science Justification and Observational Requirements – L. Keszthelyi et al.
- Titan Science with the James Webb Space Telescope (JWST) – C.A. Nixon et al.

- Observing Planetary Rings and Small Satellites with JWST: Science Justification and Observation Requirements – M.S. Tiscareno
- Cometary Science with the James Webb Space Telescope – M.S.P. Kelley et al.
- JWST observations of stellar occultations by solar system bodies and rings – P. Santos-Sanz et al.
- Physical Characterization of TNOs with JWST – A. Parker et al.

2.1 Imaging with JWST

Table 1 shows the imaging capabilities of JWST with NIRCam, NIRISS, and MIRI providing Nyquist-sampling of JWST's diffraction limited point spread function at angular resolutions of 64 mas, 130 mas, and 250 mas at wavelengths of 2, 4, and 8 microns, respectively. At wavelengths < 1.25 µm the JWST PSF is smaller than that delivered by the – comparable to the Hubble WFC3/UVIS channel at 0.5 µm (*e.g.* http://www.stsci.edu/hst/wfc3/; Dressel 2015), but the NIRCam pixels don't fully sample the PSF. Table 3 presents the spatial resolution for imaging and IFU spectra of the planets and Pluto. Saturation limits will affect most bright targets in the solar system with the full arrays, however, sub-arrays have been implemented and tested for such observations (see Norwood et al., this issue). Small bodies, rings, and fainter satellites are ideal candidates for imaging with JWST (see Keszthelyi et al, Tiscareno et al, Kelley et al., Parker et al., and Nixon et al. – this issue). As noted by those authors, many ground-breaking observations of solar system targets that have defied characterization using prior observatories will be possible using modest exposure times (up to ~1000 seconds) using JWST instrumentation. Occultations will also benefit from the imagers on JWST for observations of small bodies in the outer solar system, studies of rings, and detections of new faint bodies or rings (see Santos-Sanz et al., this issue).

2.2 Molecular Spectroscopy with JWST

Many important molecules (e.g. $H_2O$, HDO, CO, $CO_2$, $CH_4$), ices, and minerals have strong features in the JWST wavelength range; for example see Tielens (2013). Table 2 summarizes the spectral capabilities for all the instruments, with resolving power ranging from 100-2800. Solar system observations at these wavelengths, spectral resolution, and sensitivity are a major advance for compositional studies of a significant number of targets; including Mars, NEOs, TNOs, asteroids, comets, giant planets, satellites, and rings. These are all highlighted in the articles of this special issue. An example of the improvement to a previous mission to two comets that covered near infrared wavelengths is shown in Figure 3. The spectra obtained towards Comet 9P/Temple 1 with the NASA Deep Impact mission is shown compared to a simulated spectra of JWST with high spectral resolution (A'Hearn et al. 2005; Villanueva et al. 2011).

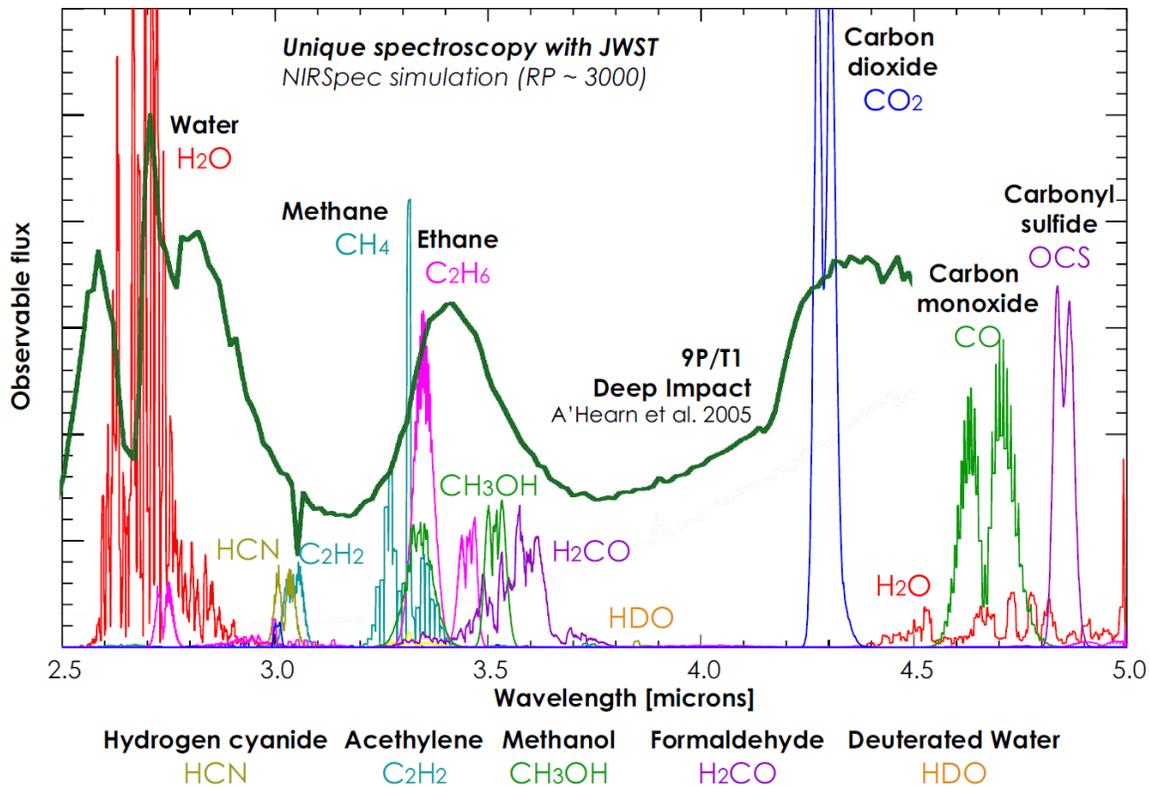

**Figure 3.** Simulated spectra of NIRSpec (RP ~ 3000) of major cometary constituents (molecular emissions arbitrarily scaled, Villanueva et al. 2011) compared to the NASA Deep Impact spectra towards comet 9P/Tempel 1 (A'Hearn et al. 2005). Significant details on composition and other physical parameters will be obtained with JWST towards multiple comets during the mission lifetime.

2.3 Comparing our Solar System to other debris disks and exoplanets

Our solar system offers the only viable possibility for characterizing objects that must be the parent bodies of the dust seen in numerous extra-solar debris disks (namely the asteroids, comets, and Kuiper Belt objects), though the effects of space weathering on small bodies may further complicate this connection and should be considered. JWST will have the sensitivity to obtain near-IR colors of any Kuiper belt object (KBO) known today with NIRCam. Much more detailed characterization can be carried out for a large subset of the known KBOs spectroscopically with NIRSpec and MIRI, and in the hermal IR with MIRI. Such observations can provide a fairly comprehensive survey of present day surface compositions of KBOs, as well as their physical properties (albedo, diameter, thermal inertia and, for the binary systems, bulk density). An extensive catalog of these intrinsic properties could finally provide solid additional insights into the dynamical processes that apparently sculpted the outer Solar System dynamical architecture (e.g. the Nice and Grand Tack models; Gomes et al. 2005; Morbidelli et al. 2005; Walsh et al. 2011), and thereby inform our thinking about dynamical processes which may operate in exo-solar planetary systems.

Our own giant planets fill a similar role in comparison to exoplanets, particularly transiting exoplanets. Studies of those distant objects are particularly expensive given current capabilities, and provide spatially and temporally limited pictures of what exoplanets are really like. Studies of the Solar System giant planets provide vastly greater detail in any given snapshot observation, and span timescales that exceed or are comparable to seasonal timescales. JWST will provide powerful new capabilities for continuing our synoptic studies of the giant planets, from multi-spectral imaging of their disks over time, to high spatial- and spectral-resolution studies of their atmospheres at wavelengths that have not previously been accessible. The richness and variability of phenomena in their atmospheres will continue to inform our thinking about exoplanets, which is also about to undergo significant advances due to the contributions JWST will make.

2.4 Bridging the Gap for Planetary Missions in the Outer Solar System

To date there are only a limited number of new science missions on the horizon for the outer solar system. Juice, the ESA Cosmic Vision Mission recently approved to start its implementation phase, is due to launch in 2022 and begin science operations in 2030. NASA recently has solicited for Discovery Class Missions, which will be selected for a launch readiness date near the end of 2021. The lack of directed missions to the outer solar system and access to wavelengths unattainable from the ground may significantly restrict the science from these bodies for a number of years. Access to facilities/observatories at multiple wavelengths for time critical events (such as storms, plumes, eruptions, impact events, etc) are essential for understanding atmospheric dynamics, chemistry, and evolution. While JWST will not be a dedicated mission to a solar system target (e.g. *Cassini*), it will still be capable for monitoring planetary (and satellite) weather for the duration of mission (minimum 5 years, probably > 10 years), 2 intervals of about 3 months each year, as well as conduct target of opportunity and time critical observations with response times of ~48 hours. The launch of JWST in 2018 will help bridge the end of the *Cassini* and *Juno* missions in 2017 to the next Discovery (if outer solar system target) or Juice (see Figure 4). Additionally, New Horizons may have an extended mission to a KBO, approaching its first target in early 2019.

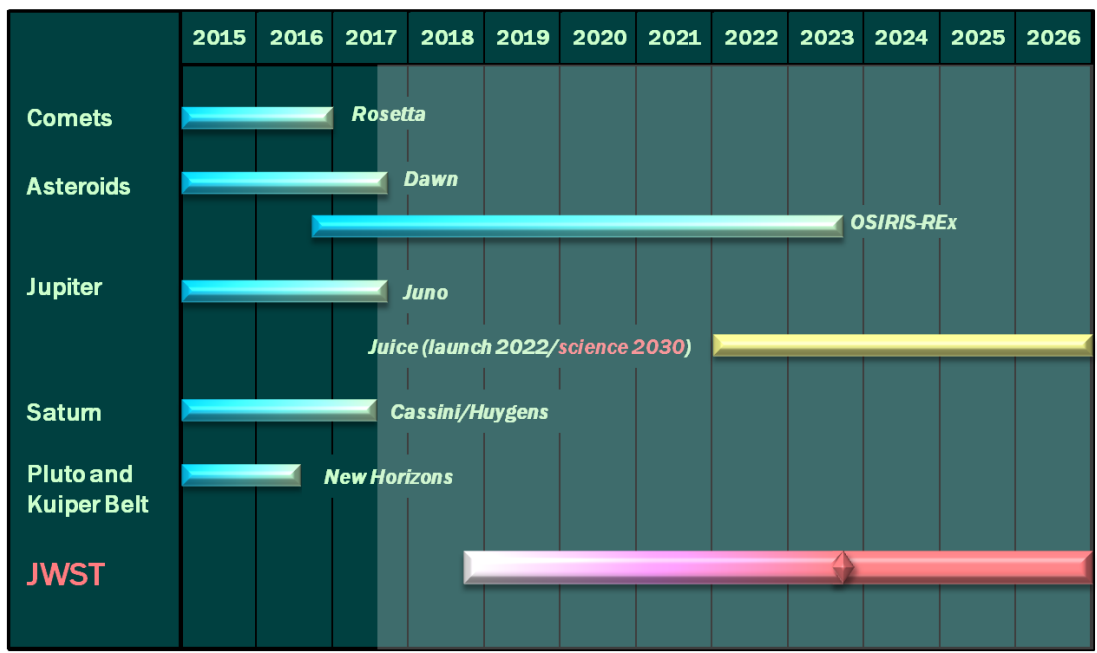

**Figure 4.** Summary of current/planned solar system missions (beyond Mars). Beyond Cassini and Juno, the next planned mission will be ESA's Juice mission due to launch in 2022 and begin science in 2030 (highlighted in the figure). The New Horizons KBO encounter/epoch is also shown here in yellow (NOTE: this phase of the mission is pending), indicating when the first target flyby would occur in Jan 2019. JWST is the only mission on the horizon that bridges that gap.

### 3. Mission timeline

JWST science observations will, for the first 2.5 years, fall into Guaranteed Time Observer (GTO) and General Observer (GO) categories. Director's Discretionary Time will also be available, and the project is considering large-program possibilities as well. GTO target lists will be finalized prior to the GO-1 call for proposals (expected to be in late 2017). GO-1 proposals will be due in early 2018. Science operations, including execution of GO-1 programs, are expected to commence in April, 2019. Mission lifetime goal is 10 years. The fraction of time allocated for Solar System proposals will approximately reflect the fraction of the total available time requested for those proposals. The selection process is expected to be highly competitive. Analysis funding will be made available to successful US-based proposers.

The Astronomer's Proposal Tool (APT, developed for the Hubble Space Telescope) is being expanded to support planning and submission of JWST proposals. Some additional APT enhancements related to planning Solar System observations will be available by GO-2 call for proposals.

More information and details about JWST, observatory and instrument capabilities, and Solar System science with JWST can be found at:
**jwst.nasa.gov/faq_solarsystem.html**
**www.stsci.edu/jwst/science/solar-system**
Instrument pocket-guides, a JWST primer, and other materials are available from:
**www.stsci.edu/jwst/science/doc-archive**
A prototype exposure-time calculator for some instrument modes is here:
**jwstetc.stsci.edu/etc**
PSF modeling software and a PSF library can be found here:
**http://www.stsci.edu/jwst/software/webbpsf**

**Tables**

**Table 1**. JWST Imaging Modes

| Mode | Instrument | Wavelength (microns) | Pixel Scale (arcsec) | Field of View* |
|---|---|---|---|---|
| Imaging | NIRCam* | 0.6 – 2.3 | 0.032 | 2.2 x 2.2' |
| | NIRCam* | 2.4 – 5.0 | 0.065 | 2.2 x 2.2' |
| | NIRISS | 0.9 – 5.0 | 0.065 | 2.2 x 2.2' |
| | MIRI* | 5.0 – 28 | 0.11 | 1.23 x 1.88' |
| Aperture Mask Interferometry | NIRISS | 3.8 – 4.8 | 0.065 | 5.1 x 5.1' |
| Coronography | NIRCam | 0.6 – 2.3 | 0.032 | 20 x 20'' |
| | NIRCam | 2.4 – 5.0 | 0.065 | 20 x 20'' |
| | MIRI | 10.65 | 0.11 | 24 x 24'' |
| | MIRI | 11.4 | 0.11 | 24 x 24'' |
| | MIRI | 15.5 | 0.11 | 24 x 24'' |
| | MIRI | 23 | 0.11 | 30 x 30'' |

* MIRI and NIRCam provide sub-array imaging to facilitate observations of bright objects, and coronagraphic imaging for the study of extra-Solar planetary systems. NIRCam has 2 modules, giving a total field of view of 2.2' x 4.4' when both are used. NIRISS AMI employs a 80 x 80 subarray.

**Table 2.** JWST Spectroscopy Modes

| Mode | Instrument | Wavelength (microns) | Resolving Power ($\lambda/\Delta\lambda$) | Field of View |
|---|---|---|---|---|
| Slitless Spectroscopy | NIRISS | 1.0 – 2.5 | 150 | 2.2' x 2.2' |
| | NIRISS | 0.6 – 2.5 | 700 | Special Mode[*] |
| | NIRCam | 2.4 – 5.0 | 2000 | 2.2' x 2.2' |
| Multi-Object Spectroscopy | NIRSpec | 0.6 – 5.0 | 100, 1000, 2700 | 3.4' x 3.4' <br> 0.2 x 0.5'' [†] |
| Single Slit Spectroscopy | NIRSpec | 0.6 – 5.0 | 100, 1000, 2700 | slit widths <br> 0.4'' x 3.8'' <br> 0.2'' x 3.3'' <br> 1.6'' x 1.6'' |
| | MIRI | 5.0 – ~14.0 | ~100 at 7.5 microns | 0.6'' x 5.5'' slit |
| Integral Field Spectroscopy | NIRSpec | 0.6 – 5.0 | 100, 1000, 2700 | 3.0'' x 3.0'' |
| | MIRI | 5.0 – 7.7 | 3500 | 3.0'' x 3.9'' |
| | MIRI | 7.7 – 11.9 | 2800 | 3.5'' x 4.4'' |
| | MIRI | 11.9 – 18.3 | 2700 | 5.2'' x 6.2'' |
| | MIRI | 18.3 – 28.8 | 2200 | 6.7'' x 7.7'' |

[*]This mode is specific to bright point source targets (e.g. exoplanets). See the NIRISS (instrument page for details (http://www.stsci.edu/jwst/instruments/niriss/science-with-niriss).
[†]Any configuration of 0.2" x 0.5" micro-shutters (365 (dispersion) x 171 (spatial) shutters per quadrant).

**Table 3.** Spatial Resolution for Imaging and IFU Spectra

| Object | Size (") | Size (km) | 2 µm PSF | NIRSpec IFU | MIRI IFU* |
|---|---|---|---|---|---|
| | | | # Resolution Elements | | |
| **Mars** | 7 | 6.8e3 | 100 | 70 | 40 |
| **Jupiter** | 37 | 1.4e5 | 530 | 370 | 200 |
| **Saturn** | 17 | 1.2e5 | 245 | 170 | 94 |
| **Uranus** | 3.5 | 5.1e4 | 50 | 34 | 19 |
| **Neptune** | 2.2 | 5.0e4 | 31 | 22 | 12 |
| **Pluto** | 0.1 | 2.4e3 | 2 | 1 | 0.6 |

*at 6.4 micron.